\documentstyle[prl,aps,preprint,epsf]{revtex}

\begin{document}


\renewcommand{\thefootnote}{\fnsymbol{footnote}}

\begin{flushright}
  \begin{tabular}{l}
  NORDITA--96--32--P\\
  JUPD--9612 \\
  MIT--CTP--2537\\
  hep-ph/9605439\\
  May 1996
  \end{tabular}
\end{flushright}
  \vskip0.5cm

\begin{center}
{\Large\bf
$Q^2$ evolution of chiral-odd twist-3 distributions $h_L(x,Q^2)$
and $e(x, Q^2)$ in large-$N_c$ QCD
}\\
\vspace{1cm}
{\sc I.I.Balitsky}${}^1$, {\sc V.M.Braun}${}^{2}$,
 {\sc Y.Koike}${}^3$ and {\sc K.Tanaka}${}^4$
\\[0.3cm]
\vspace*{0.1cm} ${}^1$ {\it Center for Theoretical Physics, MIT,
Cambridge, MA 02139, U.S.A}\\[0.3cm]
\vspace*{0.1cm} ${}^2${\it NORDITA, Blegdamsvej 17, DK-2100 Copenhagen,
Denmark}
\\[0.3cm]
\vspace*{0.1cm} ${}^3$ {\it Dept. of Physics, Niigata University,
Niigata 950--21, Japan}
\\[0.3cm]
\vspace*{0.1cm} ${}^4$ {\it Dept. of Physics, Juntendo University,
Inba-gun, Chiba 270-16, Japan}
\\[1cm]

{\bf Abstract}\\[0.3cm]
\end{center}
  We prove that the twist-3 chiral-odd parton distributions 
  obey simple Gribov-Lipatov-Altarelli-Parisi
  evolution equations in the limit $N_c\to\infty$ 
  and give analytic
  results for the corresponding anomalous dimensions. To this end we
  introduce an  evolution equation for the corresponding
  three-particle twist-3 parton correlation functions and find an exact
  analytic solution.
  For large values of $n$ (operator dimension) we are further able to 
  collect all corrections subleading in $N_c$, so our final 
  results are valid  to $O(1/N_c^2\cdot \ln(n)/n)$ accuracy.
\\


\newpage
\renewcommand{\thefootnote}{\arabic{footnote}}
\setcounter{footnote}{0}


The increasing precision of experimental data from LEP, HERA and the TEVATRON
requires understanding of higher twist corrections induced by 
correlations of partons in the colliding (produced) hadrons. 
The twist-3 parton distributions play a distinguished role in spin physics,
where they can be measured as leading effects responsible for certain 
asymmetries. In particular, the E143 collaboration has reported the 
measurement of the twist-3 contribution of the polarized structure function
$g_2(x,Q^2)$\cite{g2}, thus providing with 
the first experimental test of quark-gluon 
correlations in the nucleon. The chiral-odd parton distribution 
$h_L(x, Q^{2})$ 
is expected to be measurable in the Drell-Yan production\cite{RS,JJ}.
    
The $Q^2$ evolution of twist-3 distributions is usually believed to be 
quite sophisticated due to mixing with quark-antiquark-gluon operators,
the number of which increases with spin (moment of the structure function).
For the case of the chiral-even twist-3 flavor-nonsinglet 
distribution $g_2(x,Q^2)$ it has been observed 
by Ali-Braun-Hiller\cite{ABH}
that the operators involving gluons effectively decouple from the 
evolution equation in the large $N_c$ limit, that is neglecting $O(1/N_c^2)$
corrections \cite{remark2}. Consequently, 
$g_2(x,Q^2)$ obeys a simple GLAP evolution equation
and the corresponding anomalous dimension is known in analytic form\cite{ABH}.
The statement holds true with full account for effects
subleading in $N_c$ but for large moments $n$. 
Thus the claimed accuracy is 
in fact $O(1/N_c^2 \cdot \ln(n)/n)$. The ABH evolution equation gives a 
guide to the expected small x and large x behavior 
which is important for experimental extrapolations, and is
used\cite{bag} to rescale the model predictions to high values 
of $Q^2$ of the actual experiments.

Physics of this decoupling is so far not understood, and it is 
probably not related to usual simplifications of the $N_c\to\infty$ limit.
The observed phenomenon appears, however, to be quite general.
In this letter we demonstrate that  the same pattern  is
obeyed by chiral-odd parton distributions $h_L(x, Q^{2})$ 
and $e(x, Q^{2})$\cite{JJ} as well, albeit with different 
anomalous dimensions. For all practical purposes this solves the problem of  
the $Q^2$ evolution of twist-3 nonsinglet parton distributions,
since the corrections $1/N_c^2$ are small.

Following 
\cite{JJ,CS} 
we define the parton distributions in question 
as nucleon matrix elements of nonlocal light-cone operators\cite{remark1}
\begin{equation}
 \langle PS|\bar\psi(z)\psi(-z)|PS\rangle = 
             2 M \int_{-1}^1 dx e^{2i(P\cdot z)x}e(x,\mu^2)
\end{equation}
and
\begin{eqnarray}
\langle PS|\bar\psi(z)\sigma_{\mu\nu}z_\nu i\gamma_5\psi(-z)
         |PS\rangle 
&=&
   {2\over M} {S_{\perp}}_\mu (P\cdot z)
     \int_{-1}^1 dx e^{2i(P\cdot z)x} h_1(x,\mu^2) 
\nonumber\\  
&&{}-
2Mz_\mu \frac{(S\cdot z)}{(P\cdot z)}
     \int_{-1}^1 dx e^{2i(P\cdot z)x} h_L(x,\mu^2).
\end{eqnarray}
Here $z$ is a light-like vector $z^2=0$, 
$P$ and $S$ are the nucleon momentum
and spin vectors ($P^2=M^2,S^2=-M^2,P\cdot S=0$), respectively, and
${S_{\perp}}_{\mu} = S_{\mu} - P_{\mu}(S\cdot z)/(P \cdot z)
+ M^{2}z_{\mu} (S \cdot z)/(P \cdot z)^{2}$.
A thorough discussion of the parton interpretation and
of physical relevance of these distributions
can be found in Ref.~\cite{JJ}.

As it is well known, equations of motion allow to 
express the twist-3 
 quark-antiquark distributions in terms of quark-gluon
correlations. This is visualized  most  explicitly in form of the
operator identities (see Eqs.(27),(28) in \cite{BF2})
\begin{eqnarray}
 \bar\psi(z)\psi(-z) &=& \bar\psi(0)\psi(0)
 +\int_0^1 du\int_{-u}^u dt\, S(u,t,-u)\,,
\label{scalar}\\
 \bar\psi(z)\sigma_{\mu\nu}z_\nu i\gamma_5\psi(-z) &=&        
\big[\bar\psi(z)\sigma_{\mu\nu}z_\nu i\gamma_5\psi(-z)]_{\mbox{\rm twist 2}}
+i z_\mu \int_0^1 udu\int_{-u}^u tdt\,\widetilde S(u,t,-u),
\label{tensor}
\end{eqnarray}
where we have introduced shorthand notations for the nonlocal operators
\begin{eqnarray}
 S(u,t,v;\mu^2) &=& \bar \psi(u z)\sigma_{\mu\xi} 
   g G_{\nu\xi}(tz)z_\mu z_\nu \psi(vz)\,,
\nonumber\\
\widetilde S(u,t,v;\mu^2) &=& 
\bar \psi(u z)i\sigma_{\mu\xi}\gamma_5 
 g G_{\nu\xi}(tz)z_\mu z_\nu \psi(vz)\,.
\label{short}
\end{eqnarray}
The relations in (\ref{scalar}) and (\ref{tensor}) are exact up to
twist-4 corrections and neglecting operators containing total 
derivatives\cite{BF2} which are irrelevant for our present purposes.

Note contribution of the local scalar operator in Eq.~(\ref{scalar}). 
It gives rise to the nucleon $\sigma$-term and does not have a partonic 
interpretation. The remaining contributions give rise to three-particle
parton distributions describing  interference between scattering from
a coherent quark-gluon pair and from a single quark. 
In particular, one can define
\begin{eqnarray}
 \langle PS| S(u,t,-u)|PS\rangle &=&
  -8 M (p\cdot z)^2 \int d\alpha \int d\beta
 \, e^{i(p\cdot z)[\alpha(u-t)+\beta(u+t)]}  D_g(\alpha,\beta),
\label{distribution}
\end{eqnarray}
and a similar quantity $\widetilde{D}_g(\alpha,\beta)$ 
for $\widetilde{S}$.
The support properties of $D_g(\alpha,\beta),\widetilde D_g(\alpha,\beta)$ 
and their parton interpretation 
are discussed in Ref.~\cite{J83}.
The
variables $\alpha$, $\beta$ and $\alpha - \beta$ have physical meaning
of the momentum fractions carried 
by the antiquark-, quark- and gluon-partons,
respectively, and 
$D_g,\widetilde D_g$ vanish unless 
$|\alpha|<1$, $|\beta|<1$, and $|\alpha - \beta|<1$. 
Combining (\ref{scalar}), (\ref{tensor}) and (\ref{distribution})
one can express the twist-3 structure functions $h_L(x)$ and $e(x)$
in terms of a certain integral of these quark-antiquark-gluon 
distributions\cite{BBKT2}.
Note that only such integral (over the gluon momentum) 
is potentially measurable
in inclusive reactions like DIS or Drell-Yan processes.

The $Q^2$ dependence of the twist-3 distributions is governed by the
renormalization group (RG) equation for the corresponding nonlocal operators
$S,\widetilde S$. 
To leading logarithmic accuracy the 
evolution of $S$ and $\widetilde S$ is the same; hence we
drop the ``tilde'' in what follows.

We find it convenient to use a general approach of \cite{BB} to write
RG equations directly for the nonlocal operators. 
To this end we  introduce the Mellin transformed operators\cite{BB,ABH}
\begin{equation}
     S(u,t,v)=
     \frac{1}{2\pi i}\int^{1/2+i\infty}_{1/2-i\infty}\!\!dj\,
      (u-v)^{j-2}S(j,\xi), \hspace{0.5cm}
      \xi=\frac{u+v-2t}{u-v}.
                             \label{mellin}
\end{equation}
Here $j$ is the complex angular momentum; operators with different $j$ 
do not mix with each other. The Mellin transformed operators satisfy 
the RG equation \cite{BB} 
\begin{equation}
     \left(\mu \frac{\partial}{\partial\mu} 
           + \beta(g)\frac{\partial}{\partial g}\right)
      S(j,\xi;\mu)= -\frac{\alpha_{s}}{2\pi}
   \int_{-1}^{1}\!\!d\eta \,K_{j}(\xi,\eta)
 S(j,\eta;\mu),
                             \label{RG}
\end{equation}
where the kernel $K_{j}(\xi,\eta)$ is subject to an explicit calculation.
Neglecting all contributions which are down by $1/N_c^2$ we get 
\begin{eqnarray}
       \lefteqn{\frac{1}{N_c} K_{j}(\xi,\eta)=
                        \frac{5}{2}\delta(\xi-\eta)-
                          \frac{\theta(\xi-\eta)}{\xi-\eta}
                          \frac{1+\eta}{1+\xi}
                         \left[ \frac{1+\eta}{1+\xi}+
                          \left( \frac{1-\xi}{1-\eta}\right)^{j}\right]}
                     \nonumber\\
      &&\hspace*{2cm}\mbox{}
 +\delta(\xi-\eta)\int^{\xi}_{-1} \!\!\frac{d\eta\prime}{\xi-\eta\prime}
                          \frac{1+\eta\prime}{1+\xi}
             \left[1+\left( \frac{1-\xi}{1-\eta\prime}\right)^{2}\right]
                     \nonumber\\
      &&\hspace*{2cm}\mbox{}
                      -  \frac{\theta(\eta-\xi)}{\eta-\xi}
                          \frac{1-\eta}{1-\xi}
                         \left[ \frac{1-\eta}{1-\xi}+
                           \left( \frac{1+\xi}{1+\eta}\right)^{j}\right]
                     \nonumber\\
      &&\hspace*{2cm}\mbox{}
  +\delta(\eta-\xi)\int_{\xi}^{1}\!\!\frac{d\eta\prime}{\eta\prime-\xi}
                          \frac{1-\eta\prime}{1-\xi}
               \left[1+\left( \frac{1+\xi}{1+\eta\prime}\right)^{2}\right]
                     \nonumber\\
      &&\hspace*{1cm}\mbox{}
                           - 4\frac{\theta(\xi-\eta)}{(1+\xi)^{3}}
                            \left[\frac{2}{j}\left(1-
                  \left( \frac{1-\xi}{1-\eta}\right)^{j}\right)
                            -\frac{1-\eta}{j+1}\left(1-
      \left( \frac{1-\xi}{1-\eta}\right)^{j+1}\right)\right]
                     \nonumber\\
      &&\hspace*{1cm}\mbox{}
                           - 4\frac{\theta(\eta-\xi)}{(1-\xi)^{3}}
                            \left[\frac{2}{j}\left(1-
           \left( \frac{1+\xi}{1+\eta}\right)^{j}\right)
                            -\frac{1+\eta}{j+1}\left(1-
                  \left( \frac{1+\xi}{1+\eta}\right)^{j+1}\right)\right].
        \label{kernel}
\end{eqnarray}
The kernel turns out to be very similar to the corresponding kernel for the
evolution of chiral-even twist-3 operators in Ref.~\cite{ABH}.
Derivation of (\ref{kernel}) will be given elsewhere \cite{BBKT2}. 

 To solve the RG equation in
(\ref{RG}) we consider the  {\em conjugate} 
homogeneous equation
\begin{equation}
   \int_{-1}^{1}\!\!d\eta \,K_{j}(\eta,\xi) \phi_{j}(\eta)
=     \gamma_{j}\phi_{j}(\xi).
                             \label{r}
\end{equation}
Each eigenfunction of (\ref{RG}) corresponds to a multiplicatively
renormalizable {\em nonlocal} operator with the anomalous dimension 
$\gamma_j$: 
\begin{equation}
   \int_{-1}^{1}\!\!d\xi \,\phi_{j}(\xi) S(j,\xi;Q) = [\alpha_s(Q)/
       \alpha_s(\mu)]^{\gamma_j/b}
      \int_{-1}^{1}\!\!d\xi \,\phi_{j}(\xi) S(j,\xi;\mu);
                             \label{weight}
\end{equation}
 where $b = (11N_c-2N_f)/3$. To prove this, multiply
(\ref{RG}) by $\phi_{j}(\xi)$ and integrate over $\xi$. Using
(\ref{r}) one obtains the conventional diagonalized  RG equation 
with the solution in (\ref{weight}).

We were able to find two solutions for the 
equation in (\ref{r}) with eigenfunctions
\begin{equation}
     \phi^{+}(\xi)=1 ,\hspace{0.5cm}\phi^{-}(\xi)=\xi
                             \label{solution}
\end{equation}
(they do not depend on $j$, so we drop the subscript). 
The corresponding eigenvalues (anomalous dimensions) equal
\begin{eqnarray}
     \gamma_{j}^{+} &=&
2 N_c\left\{\psi(j+1)+\gamma_{E}-\frac{1}{4}-\frac{1}{2(j+1)}\right\}
\label{gamma1}\\
     \gamma_{j}^{-} &=&
2 N_c\left\{\psi(j+1)+\gamma_{E}-\frac{1}{4}+\frac{3}{2(j+1)}\right\}
                             \label{gamma2}
\end{eqnarray}
where $\psi(z)=\frac{d}{dz}\ln\Gamma(z)$
and $\gamma_{E}$ is the Euler constant.
Validity of Eqs. (\ref{gamma1}) and (\ref{gamma2}) can be
checked by a straightforward calculation.
By comparison with results of the numerical evaluation of anomalous 
dimensions for integer $j=n$ (see below) we conclude that our 
solutions always correspond to operators with the {\em lowest}
anomalous dimension in the spectrum. 

The superscript $\pm$ corresponds to the ``parity''
under $\xi \rightarrow - \xi$: due to the 
symmetry of the kernel one can look for separate solutions 
in the space of functions which are even (odd) with respect to 
the substitution $\xi\to -\xi$. 
{}From Eqs.(\ref{scalar}) and (\ref{tensor}), one sees that 
the relevant quantities for $e(x)$ and $h_L(x)$ are
even and odd ``$\xi$-parity'' 
pieces of the nonlocal operators.

Substituting the definition (\ref{mellin}) into (\ref{scalar}) 
and (\ref{tensor}), and changing the integration
variable $t$ to $\xi = -t/u$, 
we observe that the nonlocal 
operators in (\ref{weight}) with the particular choices of the
weight function (\ref{solution}) are precisely those which
give rise to twist-3 quark-antiquark
operators at the tree level. 
Taking the nucleon matrix elements, we get for the moments,
\begin{eqnarray}
  {\cal M}_n[e](Q) &=& L^{\gamma^+_n/b}{\cal M}_n[e](\mu)\,,
\label{moment}
\\
  {\cal M}_n[\widetilde{h}_L](Q) &=& L^{\gamma^-_n/b}
{\cal M}_n[\widetilde{h}_L](\mu),
\label{moments}
\end{eqnarray}
where ${\cal M}_n[\widetilde{h}_L]\equiv 
\int_{-1}^1 dx x^n \widetilde{h}_L(x)$,
${\cal M}_n[e]\equiv \int_{-1}^1 dx x^n e(x)$, and 
$L\equiv \alpha_s(Q)/\alpha_s(\mu)$.
$\widetilde{h}_L$ is the genuine twist-3 contribution to $h_{L}$,
after subtracting out the twist-2 piece \cite{JJ}.
The solutions for the evolution in (\ref{moment}), (\ref{moments}) 
present the principal result 
of our paper.

Expansion of nonlocal operators at small quark-antiquark separations
generates the series of local operators of increasing dimension.
In particular, expansion of $S(j, \xi)$ for positive integers $j = n \ge 2$
generates the local operators $\theta_{nk}=
\bar\psi (\stackrel{\leftarrow}{D}\cdot z)^{k-2}\sigma  G
(z \cdot \stackrel{\rightarrow}{D})^{n-k}\psi$
($k = 2, 3, \cdots$).
The substitution $\xi \rightarrow - \xi$ for the 
nonlocal operators corresponds to $k \rightarrow n - k + 2$
for the local operators.
{}From the kernel (\ref{kernel}) one can calculate the mixing matrix for
the local operators even as well as odd under 
$k \rightarrow n - k + 2$.
One can check that the result for the odd case
coincides with Eqs. (3.14)-(3.16) in Ref.\cite{KT}
to the stated $1/N_c^2$ accuracy. In particular,
neglecting the $1/N_{c}^{2}$ terms in the 
mixing matrix of \cite{KT} we obtain the evolution equation 
(\ref{moments}) with
\begin{equation}
{\cal M}_{n} [\widetilde{h}_{L}](\mu) =
\sum_{k = 2}^{[\frac{n+1}{2}]}\left( 1 - \frac{2 k}{n + 2} \right)
b_{n, k}(\mu).
\label{chk}
\end{equation}
where $b_{n, k}(\mu)$ ($k = 2, \cdots, [(n+1)/2]$) 
are reduced matrix 
elements of the independent 
quark-antiquark-gluon operators in the notation of \cite{JJ,KT}.
This is the consequence of the fact that in this limit
we have 
\begin{equation}
\sum_{k = 2}^{\left[ \frac{n+1}{2}\right]} \left(1 - \frac{2k}{n+2}
\right) X_{kl} = - \left(1 - \frac{2l}{n+2} \right) \gamma_{n}^{-},
\end{equation}
where $X_{kl}$ is the mixing matrix 
in the notation of \cite{KT} 
as 
\begin{equation}
\left(\mu \frac{\partial}{\partial \mu}
+ \beta(g) \frac{\partial}{\partial g}\right) b_{n,k}(\mu)
= \frac{\alpha_{s}}{2 \pi}\sum_{l = 2}^{\left[ \frac{n+1}{2}
\right]} X_{kl} b_{n,l}(\mu).
\end{equation}
The coefficients $\left( 1 - 2k/(n+2) \right)$
correspond to $\phi^{-}(\xi)$ of (\ref{solution})
in the nonlocal operator language.
Equation (\ref{chk}) is precisely the operator giving $\widetilde{h}_L$
at the tree level. All operators
with higher anomalous dimensions decouple 
 from the $Q^{2}$ evolution in the $N_{c}\rightarrow \infty$ limit.

The complete spectrum of anomalous dimensions in the $N_c\to\infty$ limit
obtained by the numerical diagonalization of the mixing
matrix in \cite{KT} is shown in Fig.~1, together with our analytic 
solution for the lowest eigenvalue.

To illustrate numerical accuracy of the leading-$N_c$ approximation,
consider the exact result \cite{KT} (including $1/N_c^2$ corrections) 
for the evolution of the $n=5$ moment of $h_L$, which is the lowest moment
in which mixing appears: 
\begin{eqnarray}
{\cal M}_{5}[\widetilde{h}_{L}](Q)
&=& \left[ 0.416 b_{5,2}(\mu) + 0.193 b_{5,3}(\mu) \right]
L^{12.91/b}
\nonumber \\
&+& \left[ 0.013 b_{5,2}(\mu) - 0.050 b_{5,3}(\mu) \right]
L^{18.05/b}.
\label{diag}
\end{eqnarray}
This is reduced in the large $N_c$ limit to
\begin{equation}
 {\cal M}_{5}[\widetilde{h}_{L}](Q) = \left[
\frac{3}{7}b_{5,2}(\mu) + \frac{1}{7} b_{5,3}(\mu) \right]
L^{13.7/b}.
\label{diag1}
\end{equation}
One observes: (i) the lowest anomalous dimension 
and the coefficients in front of the two terms in this contribution  
are close to their exact values; (ii) the contribution of the 
operator with the higher anomalous dimension is small ($\sim 1/N_{c}^{2}$).
It is the latter observation which is crucial for the phenomenological 
importance of our results, since it means that description of each 
moment of the twist-3 distribution requires one single nonperturbative
parameter.
A similar check for $e(x)$ [i.e., (18)-(21)] has also been done in
\cite{kn}.

We can make this point even stronger, by observing that admixture of 
operators with higher anomalous dimensions is suppressed at large $n$
for arbitrary values of $N_c$.
The full evolution equation with account of all $1/N_c^2$ terms is complicated
and will be given elsewhere \cite{BBKT2}. However, it
 simplifies drastically in the limit $j\to\infty$ and
coincides with the large-$j$ evolution equation considered in
\cite{ABH}. We get
\begin{eqnarray}
     \gamma_{j}\phi(\xi)&=&
   4C_f\left[\ln j +\gamma_{E}-3/4\right]\phi(\xi)
    +N_c \phi(\xi) \nonumber\\
               &+&N_c\!\int^{\xi}_{-1}\!\!d\eta
                \left( \frac{1-\xi}{1-\eta}\right)^{2}
                   \frac{\phi(\xi)-\phi(\eta)}{\xi-\eta}
                       +  N_c\!\int_{\xi}^{1}\!\!d\eta
                \left( \frac{1+\xi}{1+\eta}\right)^{2}
                   \frac{\phi(\xi)-\phi(\eta)}{\eta-\xi},
        \label{w}
\end{eqnarray}
with $C_f = (N_{c}^{2} - 1)/2 N_{c}$.
Comparing to the large-$j$ limit of the kernel in (\ref{kernel}), the 
only difference is in the replacement $2N_c \to 4C_f$ in the first term.
Thus, the functions in (\ref{solution}) still provide the solution,
and the anomalous dimensions are shifted by
\begin{equation}
  \gamma_j^\pm \to \gamma_j^\pm +(4C_f-2 N_c)[\ln j+\gamma_{E}-3/4].
\end{equation}
With this modification of the anomalous dimensions,
the results in (\ref{moments}) are valid to the $O(1/N_c^2\cdot \ln(n)/n)$
accuracy.

To summarize,  solutions in the present paper
provide a powerful framework both in confronting with 
experimental data and for the model-building. From a general
point of view, they are interesting as providing with 
an example of an interacting
three-particle system in which one can find an exact energy 
of the lowest state. For phenomenology, main lesson is that inclusive
measurements of twist-three distributions are complete (to our accuracy)
in the sense that knowledge of the distribution at one value of $Q_0^2$ 
is enough to predict its value at arbitrary $Q^2$, in the spirit
of GLAP evolution equation. This allows to  
relate results of different 
experiments to each other, and to compare with model calculations which 
typically are given at a very low scale. 

As shown above, the
$1/N_c^2$ corrections are not large for $n=5$ and 
further decrease as $\ln(n)/n$ at large $n$. Thus,  
the  only possibility for sizeable $1/N_c^2$ effects might be for
small $x$ behavior, in case the location of
the singularity in the complex $j$ plane  of the exact 
 anomalous dimension  
happens to be on the right of the singularity of the leading 
$N_c$ result. 
A detailed study of the small x behavior goes beyond the tasks 
of this letter.

We thank the Institute for Nuclear Theory at the University of Washington
for its hospitality and the DOE for partial support during our visit which
initiated this study.





\begin{thebibliography}{99}
\bibitem{remark2}
The leading terms in the evolution equation are proportional to $N_c$
while the neglected terms are of order $1/N_c$. Thus the accuracy is
$O(1/N_c^2)$.
\bibitem{remark1}
We do not show the gauge factors connecting the
quark and the antiquark fields.
\end{thebibliography}
\end{document}